\documentclass{article}
\usepackage{amsmath,amssymb,calc,capt-of,ifthen,epstopdf,multirow}
\usepackage{ifpdf}
\usepackage[pdftex]{graphicx}
\usepackage[a4paper,hyperindex=true]{hyperref}
\usepackage{graphicx}
\usepackage[a4paper,hyperindex=true]{hyperref}

\ifpdf
\DeclareGraphicsExtensions{.jpg}
\else
\DeclareGraphicsExtensions{.ps,.eps}
\fi

\newenvironment{itemizedot}{\begin{itemize} }{\end{itemize}}
\newcommand{\tmfloatcontents}{}
\newlength{\tmfloatwidth}
\newcommand{\tmfloat}[5]{
  \renewcommand{\tmfloatcontents}{#4}
  \setlength{\tmfloatwidth}{\widthof{\tmfloatcontents}+1in}
  \ifthenelse{\equal{#2}{small}}
    {\ifthenelse{\lengthtest{\tmfloatwidth > \linewidth}}
      {\setlength{\tmfloatwidth}{\linewidth}}{}}
    {\setlength{\tmfloatwidth}{\linewidth}}
  \begin{minipage}[#1]{\tmfloatwidth}
    \begin{center}
      \tmfloatcontents
      \captionof{#3}{#5}
    \end{center}
  \end{minipage}}

\begin{document}

\title{A Generalized Theory of Varying Alpha}
\author{John D. Barrow, Sean Z.W. Lip \\
DAMTP, Centre for Mathematical Sciences,\\
Cambridge University, Wilberforce Road,\\
Cambridge CB3 0WA, UK}
\maketitle

\begin{abstract}
In this paper, we formulate a generalization of the simple
Bekenstein-Sandvik-Barrow-Magueijo (BSBM) theory of varying alpha by
allowing the coupling constant, $\omega $, for the corresponding scalar
field $\psi $ to depend on $\psi $. We focus on the situation where $\omega $
is exponential in $\psi $ and find the late-time behaviours that occur in
matter-dominated and dark-energy dominated cosmologies. We also consider the
situation when the background expansion scale factor of the universe evolves
in proportion to an arbitrary power of the cosmic time. We find the
conditions under which the fine structure `constant' increases with time, as
in the BSBM theory, and establish a cosmic no-hair behaviour for
accelerating universes. We also find the conditions under which the fine
structure `constant' can decrease with time and compare the whole family of
models with astronomical data from quasar absorption spectra.  

PACs nos: 98.80.Es, 98.80.Bp, 98.80.Cq
\end{abstract}

\begin{center}
\textbf{I. INTRODUCTION}
\end{center}

The electromagnetic fine-structure constant, $\alpha =e^{2}/\hbar c$, has
traditionally been thought of as a fundamental constant of Nature. However,
high-redshift observations of quasar absorption spectra have continued to
suggest {\cite{OBS2, OBS1, OBS3, OBS4, OBS5} }that this `constant' may
exhibit very slow temporal and spatial variation. One theory that
generalises Maxwell's theory of electromagnetism and the general theory of
relativity in order to accommodate and test these possible variations in $%
\alpha $ is the BSBM theory of Bekenstein and Sandvik, Barrow and Magueijo {%
\cite{bek, BSBM1, BSBM2}}. In this theory, space-time variations in $\alpha $
are carried by a scalar field which couples to electrically charged matter;
this is an example of what has since become known as a `chameleonic' scalar
field. These variations, in turn, contribute to the space-time geometry and
cosmological dynamics. Such a theoretical development is important because
almost all published observational bounds on variations in $\alpha $ derive
from simply allowing $\alpha $ to become a variable in the conventional
equations of physics in which it is strictly a constant. The BSBM theory
provides a self-consistent varying-$\alpha $ theory in the way that
Brans-Dicke theory \cite{BD} describes a varying Newtonian gravitation
`constant', $G$, via a covariantly conserved Brans-Dicke scalar field $\phi
_{BD}\sim G^{-1}$.

In the BSBM theory, the quantities $c$ and $\hbar $ are taken to be
constants, and variations in $\alpha $ are ascribed to changes in $e$, the
electron charge. We write $e=e_{0}e^{\psi }$, where $\psi \equiv \psi
(x^{\mu })$ is a dimensionless scalar field, and $e_{0}$ is the present
value of $e$. The BSBM Lagrangian is {\cite{BSBM1}} 
\begin{equation}
\mathcal{L}=\sqrt{-g}(\mathcal{L}_{g}+\mathcal{L}_{mat}+\mathcal{L}_{\psi }+%
\mathcal{L}_{em}e^{-2\psi })
\end{equation}%
where $\mathcal{L}_{g}=\frac{R}{16\pi G}$, $\mathcal{L}_{\psi }=-\frac{%
\omega }{2}\partial _{\mu }\psi \partial ^{\mu }\psi $ and $\mathcal{L}%
_{em}=-\frac{1}{4}f_{\mu \nu }f^{\mu \nu }$. Here, $R$ is the Ricci
curvature scalar, $\omega $ is a coupling constant, and we have defined an
auxiliary gauge potential $a_{\mu }=e^{\psi }A_{\mu }$ with corresponding
field tensor $f_{\mu \nu }=e^{\psi }F_{\mu \nu }=\partial _{\mu }a_{\nu
}-\partial _{\nu }a_{\mu }$, so that the covariant derivative takes its
usual form $D_{\mu }=\partial _{\mu }+ie_{0}a_{\mu }$. The quantity $%
\mathcal{L}_{em}$ is usually parametrized by the ratio $\zeta =\text{$%
\mathcal{L}_{em}$}/\rho $, where $\rho $ is the total baryon energy density.
This ratio, $\zeta $, describes the fraction of non-relativistic matter in
the universe that contributes to $\text{$\mathcal{L}$}_{\text{em}}$. The
cosmological value of $\zeta $ (denoted $\zeta _{m}$) has to take into
account non-baryonic matter, and thus depends strongly on the nature of dark
matter, the nature and constituents of which are currently uncertain. In
general, therefore, $\zeta _{m}$ is determined by the relative role of
magnetic, $B,$ and electric, $E,$ fields in the dominant cold dark matter
content of the universe, and $\zeta _{m}=(E^{2}-B^{2})/(E^{2}+B^{2})$ can
take any value between $-1$ and $1$ \cite{BSBM1},\cite{BSBM2}; also, during
the radiation era, $\langle\zeta _{m}\rangle$ $=0$. In the situations where $%
\zeta _{m}<0,$ a slow logarithmic increase in $\alpha $ occurs during the
matter-dominated era but $\alpha $ then tends to a constant when dark energy
causes the expansion of the universe to accelerate \cite{BSBM2}. Therefore,
if $\alpha $ is found to decrease with time during the matter-dominated era,
the simple BSBM model with $\zeta _{m}<0$ would be inconsistent with this
variation.

Observational searches for varying $\alpha $ at cosmological redshifts have
used high-precision studies of quasar spectra coupled with many-body
calculations of the effects of small changes to $\alpha $ on relativistic
corrections to atomic transition frequencies \cite{lines}. The most recent
analysis by Webb et al {\cite{OBS5} }of a new set of observations of quasar
spectra from the VLT gives the following best fits for $\Delta \alpha
/\alpha \equiv (\alpha (z)-\alpha _{0})/\alpha _{0},$where $\alpha _{0}$
denotes the present value of $\alpha $ and $z$ is the redshift of the
absorption lines:

\ \ \ \ \ \ \ \ \ \ \ \ \ \ \ \ \ \ $\Delta \alpha / \alpha = (- 0.06 \pm
0.16) \times 10^{- 5}$ for $z < 1.8$,

\ \ \ \ \ \ \ \ \ \ \ \ \ \ \ \ \ \ $\Delta \alpha / \alpha = (0.61 \pm
0.20) \times 10^{- 5}$ for $z > 1.8$.

For comparison, an earlier analysis of data from the Keck Telescope gave {\ {%
\cite{OBS2}}, \cite{OBS1}}, {\cite{OBS4}}:

\ \ \ \ \ \ \ \ \ \ \ \ \ \ \ \ \ \ $\Delta \alpha / \alpha = (- 0.54 \pm
0.12) \times 10^{- 5}$ for $z < 1.8$,

\ \ \ \ \ \ \ \ \ \ \ \ \ \ \ \ \ \ $\Delta \alpha /\alpha =(-0.74\pm
0.17)\times 10^{-5}$ for $z>1.8$.

All the errors quoted are $1\sigma $ errors. Note that, for redshifts $z>1.8$%
, the sense in which $\alpha (z)$ varies is different for these two sets of
observations. The two samples were obtained from telescopes in different
hemispheres of the Earth, which has led to a proposal in \cite{OBS5} that
there may be a\ strong spatial (dipolar) dependence in the variation of $%
\alpha $. However, the sense of the overall temporal variation is unclear,
and it is conceivable that $\alpha $ could increase, decrease or remain
constant with time during the matter era. Further evidence that $\alpha $
may increase during part of the matter-dominated era can be found from
observations based on single ion differential measurements, which place a
constraint of $\Delta \alpha /\alpha =(0.566\pm 0.267)\times 10^{-5}$ at $%
z=1.84$ {\cite{SIDAM1}}, {\cite{SIDAM2} that the authors interpret as an
upper limit because of unknown systematics}. For a summary of these results,
and others, the reader is referred to the recent theoretical reviews by
Martins \cite{mar} and Uzan {\cite{UZAN}}. However, note that the
observational results announced by Srianand et al \cite{chand}, which
appeared at first to be consistent with no significant redshift variation in 
$\alpha ,$ have been shown to have methodological problems, and a subsequent
re-analysis of the same data produced results consistent with the variations
found in {\cite{OBS2, OBS1, OBS3, OBS4}. For a discussion of some of the
observational problems, see \cite{webb}. }

In this paper, we will show how the simple BSBM theory can be generalised in
a natural manner by allowing the coupling constant $\omega $ in $\mathcal{L}%
_{\psi }$ to take on a $\psi $-dependence, so that $\omega =\omega (\psi )$.
We will show that a simple choice of the coupling function, $\omega (\psi ),$
permits $\Delta \alpha /\alpha $ to increase or decrease during the
matter-dominated era according to the value of a single parameter which
vanishes when the new theory reduces to the BSBM case (with constant $\omega 
$). However, there is still a general pattern to the late-time $\alpha (t)$
behaviour in an expanding universe, and solutions for $\alpha (t)$ generally
tend to a particular asymptotic form which is independent of the initial
conditions. We shall focus our analysis on the late-time behaviour of the
matter-dominated and $\Lambda $-dominated eras of a zero-curvature Friedmann
universe, but also provide solutions for the more general case in which the
scale factor of the universe evolves as $a(t)=t^{n}$, where $n$ is a
constant, which mimics negative curvature dominated expansion when $n=1$. We
find that, as in the original BSBM theory, this extended theory still
predicts that any variation in $\alpha $ is rapidly suppressed during the $%
\Lambda $-dominated era, thus allowing complex atoms to persist into the far
future (see \cite{barflat}). Our generalised theory also gives more scope
for spatial variations in $\alpha $ to be described because gradients in $%
\alpha $ can now be driven by gradients in $\omega (\psi )$.

In section II we set up the basic equations of the generalized BSBM theory
and explore some of their key properties. In section III we discuss the
choice of $\omega (\psi )$ and in section IV we study the cosmological
solutions in a dust-dominated flat Friedmann universe. In section V we
compare this with the behaviour in a wider range of Friedmann background
metrics. In section VI we investigate what happens to the variation of $%
\alpha $ when the universe undergoes accelerated expansion, dominated by
dark energy in the form of a positive cosmological constant. In section VII
we summarise these results and compare them with the cosmological data. 
Finally, section VIII contains some concluding discussion.

\begin{center}
\textbf{II. THE UNDERLYING MODEL AND BACKGROUND EQUATIONS}
\end{center}

Consider the modified BSBM Lagrangian 
\begin{equation}
\mathcal{L}=\sqrt{-g}(\mathcal{L}_{g}+\mathcal{L}_{\text{mat}}+\mathcal{L}%
_{\psi }+\mathcal{L}_{\text{em}}e^{-2\psi })  \label{LAG}
\end{equation}%
where $\mathcal{L}_{g}=\frac{R}{16\pi G}$, $\mathcal{L}_{\psi }=-\frac{%
\omega (\psi )}{2}\partial _{\mu }\psi \partial ^{\mu }\psi $ and $\mathcal{L%
}_{\text{em}}=-\frac{1}{4}f_{\mu \nu }f^{\mu \nu }$, where the coupling $%
\omega $ is now taken to be a function of the scalar field $\psi $. In order
to ensure that the energy density of the scalar field, $\rho =\frac{\omega
(\psi )}{2}\dot{\psi}^{2}$, is non-negative, we will impose the no-ghost
condition $\omega (\psi )\geqslant 0$. As in the BSBM theory, the fine
structure `constant' is determined \ by 
\begin{equation}
\alpha =\alpha _{0}\exp (2\psi ).  \label{alpha}
\end{equation}%
Varying (\ref{LAG}) with respect to $\psi $ gives the equation of motion for
the scalar field: 
\begin{equation}
0=\sqrt{-g}\left( \frac{\omega ^{\prime }(\psi )}{2}\partial _{\mu }\psi
\partial ^{\mu }\psi -2\mathcal{L}_{\text{em}}e^{-2\psi }+\omega (\psi
)\partial _{\mu }\partial ^{\mu }\psi +\omega (\psi )\partial ^{\mu }\psi 
\frac{\partial _{\mu }\sqrt{-g}}{\sqrt{-g}}\right) .  \label{scalar}
\end{equation}

We now specialize to the case of a spatially-flat Friedmann universe with
coordinates $(t,x^{i})$ and metric $g_{\mu \nu }=\text{diag}%
(-1,a^{2}(t),a^{2}(t),a^{2}(t))$. We will adopt a system of units in which $%
G=1$ and $c=1$. The modified Friedmann equation, obtained by varying (\ref%
{LAG}) with respect to the metric, is: 
\begin{equation}
\frac{\dot{a}^{2}}{a^{2}}=\frac{8\pi }{3}\left( \rho _{m}\left( 1+|\zeta
|e^{-2\psi }\right) +\rho _{nb}+\rho _{r}e^{-2\psi }+\rho _{\psi }+\rho
_{\Lambda }\right)   \label{FRIED}
\end{equation}%
Here, an overdot indicates a derivative with respect to the comoving proper
time $t$. In this equation, $\rho _{\Lambda }=\frac{\Lambda }{8\pi }$ is a
constant denoting the density of dark energy, $\rho _{nb}\propto a^{-3}$ is
the contribution of non-baryonic matter that is assumed to form the cold
dark matter, and $\rho _{\psi }=\frac{\omega (\psi )}{2}\dot{\psi}^{2}$.
This equation is unchanged from the corresponding equation for BSBM, since
any new contributions due to the $\psi $-dependence of $\omega $ will be of
the form $\frac{\partial \mathcal{L}}{\partial \psi }\frac{\partial \psi }{%
\partial g_{\mu \nu }}$, and we have $\frac{\partial \psi }{\partial g_{\mu
\nu }}=0$. Furthermore, (\ref{FRIED}) can be approximated by a standard
Friedmann equation at late times if the conditions 
\begin{equation}
|\zeta |e^{-2\psi }\ll 1\text{ and }\dot{\psi}^{2}\omega \ll \rho _{c}
\label{consist}
\end{equation}%
are satisfied then, where $\rho _{c}$ represents the dominant form of matter
in the universe during the epoch being considered. In this paper we will be
mainly interested in the cold dark matter and $\Lambda $-dominated eras, in
which $\rho _{c}=\rho _{m}$ and $\rho _{c}=\rho _{\Lambda }$ respectively.
(Note that the first condition in (\ref{consist}) is automatically satisfied
if $\psi $ increases to infinity at late times. However, this is not a
necessary requirement: for example, $|\zeta |$ could be small and $\psi $
could tend to a large constant.)

We will call these two conditions the {\itshape{consistency conditions}},
and will use them to check that our late-time solutions are consistent with
our assumption that the solution for the expansion scale factor, $a(t)$, is
unaffected by the small variations in $\psi $ that we are interested in
modelling. In contrast, the evolution of $\psi $ is significantly affected
by the cosmological expansion.

The evolution equation (\ref{scalar}) for $\psi $ becomes 
\begin{equation}
\ddot{\psi}+3H\dot{\psi}+\frac{\omega ^{\prime }(\psi )}{2\omega }\dot{\psi}%
^{2}=-\frac{2\zeta \rho _{m}}{\omega }e^{-2\psi }  \label{COSM}
\end{equation}%
and, after substituting $\omega ^{\prime }(\psi )\dot{\psi}=\dot{\omega}$,
we can rewrite this as: 
\begin{equation}
\left( \omega ^{\frac{1}{2}}\dot{\psi}a^{3}\right) ^{.}\equiv \frac{N}{%
\omega ^{1/2}}e^{-2\psi }  \label{MASTER}
\end{equation}%
where $N\equiv -2\zeta \rho _{m}a^{3}$ is a constant. In some sense, this
equation is just the BSBM equation in disguise. This can be seen by defining
a new time coordinate $T$ with $\frac{d}{dT}=\omega ^{1/2}\frac{d}{dt}$, so
that $T=\int \omega ^{-1/2}dt$. (Since $\omega ^{-1/2}>0$, this is a
monotonic reparametrisation.) Then (\ref{MASTER}) becomes 
\begin{equation}
\frac{d}{dT}\left( \frac{d\psi }{dT}a^{3}\right) =Ne^{-2\psi }  \label{NEWT}
\end{equation}%
which is of the same form as the corresponding BSBM equation for $\psi $,
but with a different time parameter, $T$. However, this observation is of
limited use for finding solutions, because we do not know the behaviour of $%
\psi (t)$ at the outset, and so we do not know the behaviour of $\omega
(\psi (t))$.

Inspection of the conservation equation (\ref{COSM}) reveals an important
feature of the evolution of $\psi $ (and hence $\alpha $). In the case $N>0$%
, if $\dot{\psi}$ vanishes anywhere, then $\ddot{\psi}>0$ at that turning
point, so $\psi $ can be monotonic or have a minimum, but it cannot have a
maximum. Similarly, in the case $N<0$, $\psi $ can be monotonic or have a
maximum, but it cannot have a minimum. Therefore, in either case, $\psi $
cannot oscillate, although some analyses \cite{cop} of the linearised
version of (\ref{COSM}) have been misled into deducing oscillatory behaviour
for $\alpha $.

At present, $\omega (\psi )$ is an arbitrary function. However, if the
observed behaviour of $\alpha (z)$ and $\psi (t)$ is monotonic, it should be
possible (in principle) to obtain the form of $\omega (\psi )$ if our
observational data are sufficiently accurate to allow us to fit a function
for $\alpha (z)$, and hence determine $\psi (t)$. This can be done as
follows. First, note that equation (\ref{MASTER}) can be re-written as a
first-order differential equation for $\omega (t)$: 
\begin{equation*}
\frac{\dot{\psi}a^{3}}{2}\dot{\omega}+\omega (\dot{\psi}a^{3})^{\dot{}%
}=Ne^{-2\psi }
\end{equation*}%
which has the general solution 
\begin{equation}
\omega (t)=\frac{2N}{(\dot{\psi}a^{3})^{2}}\int e^{-2\psi }\dot{\psi}a^{3}dt.
\label{SOLN}
\end{equation}%
Given an expression for $\psi (t)$, we can calculate the right-hand side of (%
\ref{SOLN}) in terms of $t$, and then write $t=t(\psi )$ to obtain an
expression for $\omega (\psi )$.

For example, in the matter era with $a\propto t^{2/3}$ and $\rho _{m}\propto
t^{-2}$, the general solution (\ref{SOLN}) allows us to solve explicitly for 
$\omega (t)$, given a specified behaviour for $\psi (t)$: 
\begin{equation*}
\omega (t)=\frac{2N}{(\dot{\psi}t^{2})^{2}}\int e^{-2\psi }\dot{\psi}t^{2}dt.
\end{equation*}%
If, for example, $\psi =A\ln t$, where $A>0$ and $A\neq 1$, then 
\begin{equation*}
\omega (t)=\frac{Nt^{-2A}}{A(1-A)}+\frac{k}{t^{2}}=\frac{Ne^{-2\psi }}{A(1-A)%
}+ke^{-2\psi /A}.
\end{equation*}%
On the other hand, if $\psi =\ln t$, then 
\begin{equation*}
\omega (t)=\frac{1}{t^{2}}\left( 2N\ln t+k\right) =e^{-2\psi }(2N\psi +k),
\end{equation*}%
where $k$ is an arbitrary constant. It is easy to check that the consistency
conditions (\ref{consist}) hold in both cases (which implies that the
approximate cosmological solution $a\sim t^{2/3}$ is valid).\emph{\ }

\begin{center}
\textbf{III. A CHOICE OF $\omega(\psi)$}
\end{center}

The coupling function $\omega (\psi )$ is a free function and its
introduction generalizes the BSBM model, in which $\omega $ was a constant.
This generalization introduces another way in which spatial inhomogeneity
can be created in $\psi $ and $\alpha $. We are interested here in
situations where $\omega (\psi )$ remains non-negative and finite for all
real values of $\psi $. We make the choice $\omega (\psi )\propto e^{\mu
\psi }$, where $\mu $ is a constant to be determined or constrained by
observation. Other forms of $\omega (\psi )$ either take negative values
(e.g. $\omega (\psi )=\psi $), blow up at the origin (e.g. $\omega (\psi
)=\psi ^{-2}$) or are more complicated than the simple form above. Thus, in
the remainder of this paper, we will restrict ourselves to considering the
form 
\begin{equation}
\omega (\psi )=\omega _{0}e^{\mu \psi },  \label{exp}
\end{equation}%
where $\omega _{0}$ and $\mu $ are constants. Note that the BSBM theory is
recovered in the $\mu \rightarrow 0$ limit.

\begin{center}
\textbf{IV. SOLUTIONS IN THE MATTER ERA}
\end{center}

We need to investigate the cosmological evolution of $\psi $ in the epoch
that includes the astronomical observations of quasars and low-redshift
information on the possible evolution of $\alpha $, so we begin our study in
the matter-dominated era. We will assume that the universe contains a
perfect `dust' fluid, with zero pressure, and that the background scale
factor is of the form $a(t)=a_{0}t^{2/3}$. We will need to check that the
consistency conditions (\ref{consist}) hold in order to ensure that we can
ignore any corrections to the scale factor evolution from the evolution of $%
\psi (t)$. We will present exact solutions for certain special cases, but
for the general behaviour we will use numerical investigations to provide
insight into the evolutionary behaviour of $\alpha $.

From (\ref{COSM}), the general evolution equation for $\psi (t)$ in a
Friedmann background with expansion scale factor $a(t)$ is 
\begin{equation}
\ddot{\psi}+\frac{\mu }{2}\dot{\psi}^{2}+3\frac{\dot{a}}{a}\dot{\psi}=\frac{%
Ca_{0}^{3}e^{-(2+\mu )\psi }}{a^{3}}  \label{evol}
\end{equation}%
where $\omega =\omega _{0}e^{\mu \psi }$ and $C=N/(\omega _{0}a_{0}^{3})$.
In the matter-dominated era, the scale factor evolves as $a(t)=a_{0}t^{2/3}$%
, so this equation reduces to 
\begin{equation}
\ddot{\psi}+\frac{\mu }{2}\dot{\psi}^{2}+\frac{2}{t}\dot{\psi}=\frac{%
Ce^{-(2+\mu )\psi }}{t^{2}}.  \label{ref3A}
\end{equation}%
The consistency conditions are 
\begin{equation*}
|\zeta |e^{-2\psi }\ll 1\text{and }e^{\mu \psi }\dot{\psi}^{2}t^{2}\ll 1\ \
\ \ \ \text{ as }t\rightarrow \infty .
\end{equation*}%
The equation (\ref{ref3A}) can be transformed into an Emden-Fowler equation
using the substitution $\psi =\frac{2}{\mu }\ln \left( \theta t^{-1}\right) $%
: 
\begin{equation*}
\text{$\ddot{\theta}=\frac{\mu C}{2}\theta ^{-\frac{4}{\mu }-1}t^{\frac{4}{%
\mu }}$}.
\end{equation*}%
Note that $\theta (t)$ must be positive, and that $\ddot{\theta}$ always has
the same sign. The consistency conditions become $\left( \theta /t\right)
^{-4/\mu }\ll 1$ and $\left( \dot{\theta}-\theta t^{-1}\right) ^{2}\ll 1$.
The second condition implies that $\theta (t)$ must grow at most as $t\ln t$
as $t\rightarrow \infty $ (and that if $\theta (t)\sim t\ln t$ then the
constant of proportionality should be very small). In applying these
conditions, however, we note that in the standard cosmology the matter era
lasts for a limited period of time before giving way to an accelerated
expansion with $p\simeq -\rho $. Thus, for example, if $\mu <0$, then a
solution with $\theta $ growing asymptotically faster than $t$ would be
acceptable, as long as $\theta (t)\leqslant \kappa t$ throughout the matter
epoch (where $\kappa $ is a small constant).

For two special cases, exact solutions can be found:

{\bfseries{Case S1:}} In the case $\mu = -2$, the governing equation is $%
\ddot{\theta} = - \frac{C \theta}{t^2}$. This equation is equidimensional
and has solutions of the form $\theta (t)=t^{k}$, where $k$ is a constant.
If $C=\frac{1}{4}$, then 
\begin{equation*}
\text{$\theta = \beta_1 \sqrt{t} + \beta_2 \sqrt{t} \ln t$}
\end{equation*}
where we require that $\beta_2 > 0$, or that $\beta_2 = 0$ and $\beta_1 > 0$%
. This solution satisfies both consistency conditions. The corresponding
solution for $\psi$ is 
\begin{equation*}
\text{$\psi = \frac{1}{2} \ln t - \ln \left( \beta_1 + \beta_2 \ln t \right)$%
} .
\end{equation*}
In this (very) special case, $\psi$ therefore increases logarithmically at
late times and the fine structure `constant' evolves as 
\begin{equation*}
\alpha \equiv \exp (2\psi )=\frac{t}{\left( \beta_{1}+\beta_{2}\ln t\right)
^{2}}\sim \frac{t}{(\ln t)^{2}}.
\end{equation*}

On the other hand, if $C<\frac{1}{4}$, then 
\begin{equation*}
\text{$\theta =\beta _{1}t^{\frac{1-\sqrt{1-4C}}{2}}+\beta _{2}t^{\frac{1+%
\sqrt{1-4C}}{2}}$}
\end{equation*}%
at late times, where, again, we require that $\beta _{2}>0$, or that $\beta
_{2}=0$ and $\beta _{1}>0$. In the former case, we require $C>0$ for the
first consistency condition to hold; in the latter case, there is no
condition on $C$. (As noted above, it is possible to have a consistent
solution in the former case when $C<0$, as long as $\beta _{2}$ is very
small.) The corresponding solution for $\psi $ is 
\begin{equation*}
\psi =\frac{1}{2}\left( 1-\sqrt{1-4C}\right) \ln t-\ln \left( \beta
_{2}+\beta _{1}t^{-\sqrt{1-4C}}\right) .
\end{equation*}%
At late times, this solution tends to $\psi =\frac{1}{2}\left( 1-\sqrt{1-4C}%
\right) \ln t-\beta _{1}t^{-\sqrt{1-4C}}$ and $\alpha $ evolves as 
\begin{equation*}
\alpha =\frac{t^{1-\sqrt{1-4C}}}{\left( \alpha _{2}+\alpha _{1}t^{-\sqrt{1-4C%
}}\right) ^{2}}\sim t^{1-\sqrt{1-4C}}.
\end{equation*}%
Note that, in this case, $\psi $ decreases at late times if and only if $C<0$%
.

Finally, if $C > \frac{1}{4}$, we obtain 
\begin{equation*}
\text{$\theta = t^{\frac{1}{2}} \left( \beta_1 \cos \left( \frac{\sqrt{4 C -
1}}{2} \ln t \right) + \beta_2 \sin \left( \frac{\sqrt{4 C - 1}}{2} \ln t
\right) \right)$.}
\end{equation*}
In this case, physical solutions do not exist because $\theta (t)$ takes
negative values.

{\bfseries{Case S2:}} In the case $\mu =-4$, the governing equation becomes $%
\text{$\ddot{\theta}=-2Ct^{-1}$}$ with the exact solution $\theta =-2Ct\ln
t+\beta _{1}t+\beta _{2}$, where $\beta _{1}$ and $\beta _{2}$ are arbitrary
constants. Since we require $\theta (t)$ to be positive (at least at late
times), this means that a solution exists only if $C<0$. The corresponding
solution for $\psi $ is 
\begin{equation*}
\psi =-\frac{1}{2}\ln \left( -2C\ln t+\beta _{1}+\frac{\beta _{2}}{t}\right)
,
\end{equation*}%
and $\alpha \sim 1/\ln t$ at late times. This solution decreases at late
times, and therefore it is necessary to impose the requirement that $-2C\ln
t_{m\Lambda }\ll 1$, where $t_{m\Lambda }$ is the time at which the matter
era ends.

\begin{figure}[tbh]
\begin{center}
\includegraphics[width=0.9\textwidth,height=0.7\textheight]{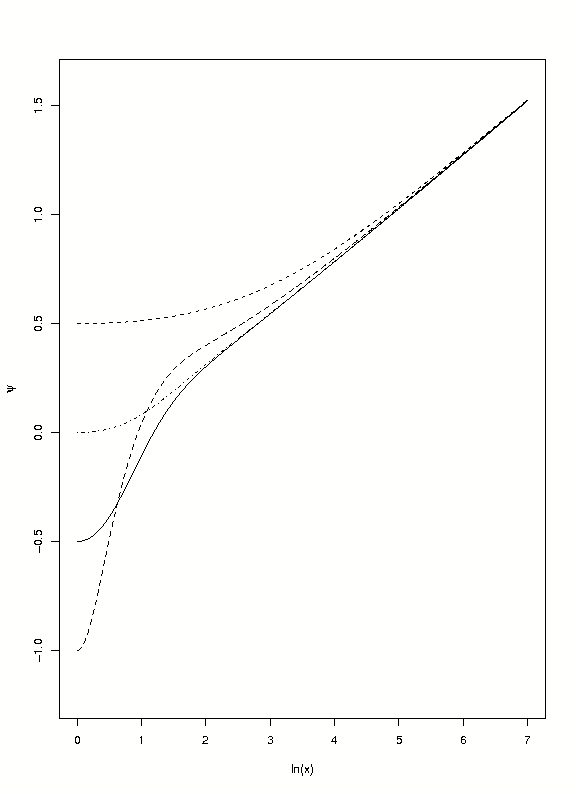}
\end{center}
\caption{A plot of four solutions of equation (\protect\ref{ref3A}) for the
various choices $\protect\psi (\ln x=0)\in \{-1,-0.5,0,0.5\}$ and $d\protect%
\psi /d(\ln x)|_{\ln x=0}=0$. Here, $x$ is defined as $\ln (t/t_{1})$, where 
$t_{1}$ is an arbitrary constant, and we have chosen (arbitrarily) the
parameters $\protect\mu =2$ and $C=0.1$. Notice that, in this case $(\protect%
\mu >-2)$, all the solutions eventually converge to a common increasing
asymptote, independently of the initial conditions.}
\label{fig:1}
\end{figure}

We now consider general values of $\mu $. We determine asymptotic forms for
the solutions and illustrate the behaviour with numerical plots of $\psi $
against $\ln (\ln t))$ for a wide range of initial conditions. Remarkably,
there are just two simple asymptotes:

{\bfseries{Case 1:}} $C>0$: Here, there are two cases depending on the value
of $\mu $. If $\mu <-2$, there is a finite-time singularity at which $\psi $%
, and thus $\alpha $, becomes unboundedly large. However, if $\mu >-2$, all
solutions tend to a common asymptote, similar to the behaviour found during
the matter era in the standard BSBM scenario described in {\cite{BSBM2}}
(Fig. \ref{fig:1}). This result does not seem to depend on the initial
conditions. The rate at which $\psi $ increases along this asymptote can be
made as gentle as desired by taking $\mu $ sufficiently large (Fig. \ref%
{fig:2}).

\begin{figure}[tbh]
\begin{center}
\includegraphics[width=0.9\textwidth,height=0.7\textheight]{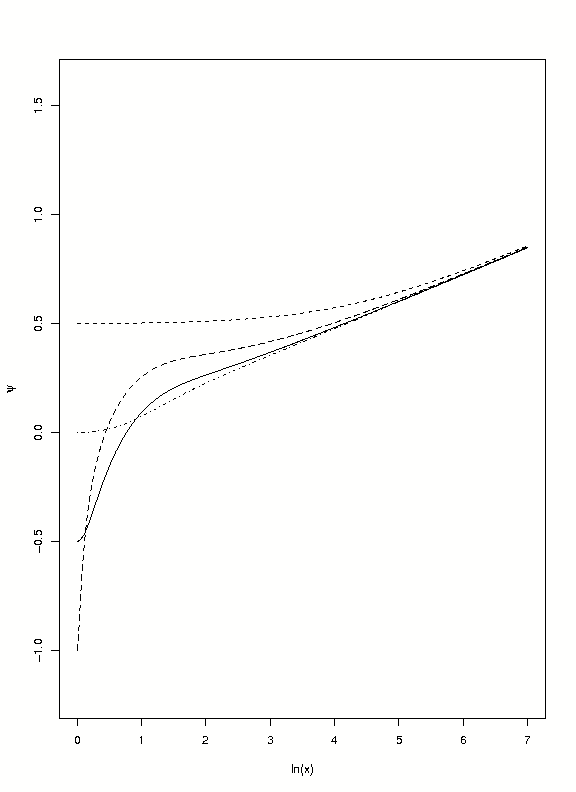}
\end{center}
\caption{As in Figure \protect\ref{fig:1}, but with $\protect\mu =6$. All
solutions still converge to a common increasing asymptote at late times.
However, the slope of the asymptote is more gentle than that in Figure 
\protect\ref{fig:1}, because the value of $\protect\mu $ has increased. The
slope of the asymptote can be made as gentle as desired by taking $\protect%
\mu $ to be sufficiently large.}
\label{fig:2}
\end{figure}

We will now obtain an explicit formula for the asymptote by showing that, at
late times, 
\begin{equation*}
\psi \rightarrow \frac{1}{2+\mu }\ln \left( (2+\mu )C\ln t\right)
\end{equation*}%
and so 
\begin{equation*}
\alpha \rightarrow (\ln t)^{2/(2+\mu )}
\end{equation*}%
Note that this is consistent with the earlier BSBM result {\cite{BSBM2},} $%
\psi \rightarrow \frac{1}{2}\ln \left( 2C\ln t\right) ,$ which is recovered
when $\mu \rightarrow 0$.

In order to extract the asymptotic solution in finer detail we will proceed
as in {\cite{BSBM2}}. Using the substitution $x=\ln (t/t_{1})$, where $t_{1}$
is an arbitrary constant, we can transform (\ref{ref3A}) to the following
equation: 
\begin{equation}
\psi ^{\prime \prime }+\psi ^{\prime }+\frac{\mu }{2}\psi ^{\prime
2}=Ce^{-(2+\mu )\psi }  \label{Puiseux}
\end{equation}%
where a prime denotes $d/dx$. We now write 
\begin{equation*}
\psi =A\ln (Bx)+\sum_{n=1}^{\infty }a_{n}x^{-n}.
\end{equation*}%
Substituting this series into (\ref{Puiseux}) gives 
\begin{eqnarray*}
Ax^{-1} &+&\left( \frac{\mu }{2}A^{2}-A-a_{1}\right)
x^{-2}+\sum_{n=2}^{\infty }\left( (n-\mu A)(n-1)a_{n-1}-na_{n}\right)
x^{-n-1} \\
&+&\frac{\mu }{2}\left( \sum_{n=1}^{\infty }na_{n}x^{-n-1}\right)
^{2}=C(Bx)^{-(2+\mu )A}\exp \left( ^{{}}-(2+\mu )\sum_{n=1}^{\infty
}a_{n}x^{-n}\right) .
\end{eqnarray*}%
In order for the leading terms (in $x^{-1}$) to match, we need $-1=-(2+\mu
)A $ and thus $A=\frac{1}{2+\mu }$. We also need $A=CB^{-(2+\mu )A}$ which
forces $B=(2+\mu )C$. This leads to 
\begin{eqnarray*}
\frac{1}{(2+\mu )x} &+&\left( \frac{\mu }{2}A^{2}-A-a_{1}\right)
x^{-2}+\sum_{n=2}^{\infty }\left( (n-\mu A)(n-1)a_{n-1}-na_{n}\right)
x^{-n-1} \\
&+&\frac{\mu }{2}\left( \sum_{n=1}^{\infty }na_{n}x^{-n-1}\right) ^{2}=\frac{%
1}{(2+\mu )x}\left( e^{-(2+\mu )\sum_{n=1}^{\infty }a_{n}x^{-n}}\right) .
\end{eqnarray*}%
The successive terms of $(a_{n})_{n=1}^{\infty }$ can now be chosen so as to
make the left-hand side vanish. This requires 
\begin{equation*}
a_{1}=-\frac{4+\mu }{2(2+\mu )^{2}},\text{ }a_{2}=-\frac{1}{4}\frac{(4+\mu
)^{2}}{(2+\mu )^{3}},\text{ etc}.
\end{equation*}%
which leads to 
\begin{equation*}
\frac{1}{(2+\mu )x}=\frac{1}{(2+\mu )x}\exp \left( -(2+\mu
)\sum_{n=1}^{\infty }a_{n}x^{-n}\right) \rightarrow \frac{1}{(2+\mu )x}
\end{equation*}%
as $x\rightarrow \infty $. Therefore, we have 
\begin{equation*}
\psi =\frac{1}{2+\mu }\ln ((2+\mu )Cx)+\sum_{n=1}^{\infty }a_{n}x^{-n}
\end{equation*}%
and 
\begin{equation}
\alpha =e^{2\psi }=\left[ (2+\mu )C\ln (t/t_{1})\right] ^{\frac{2}{2+\mu }%
}\exp \left( -\frac{4+\mu }{(2+\mu )^{2}\ln (t/t_{1})}\right)
\label{alphaseries}
\end{equation}%
to leading order.

In order for this solution to hold during the matter era, the value of $%
t_{1} $ should be less than the value of $t$ at the beginning of the
matter-dominated era.

{\bfseries{Case 2:}} $C<0$. The analysis again breaks down into two cases.
If $\mu >-2$, then at some finite future time, $\psi $ becomes unboundedly
negative, driving $\alpha $ to zero; we ignore this case with a finite-time
singularity. However, if $\mu <-2$, the behaviour mimics the $C>0,\mu >-2$
behaviour, and solutions tend to a common asymptote 
\begin{equation*}
\psi \rightarrow \frac{1}{2+\mu }\ln \left( (2+\mu )C\ln (t/t_{1})\right)
\end{equation*}%
where, again, $t_{1}$ is an arbitrary constant. The analysis is similar to
the above, but now $\psi $ and $\alpha $ both decrease with time (Figs. \ref%
{fig:3} and \ref{fig:4}).

\begin{figure}[tbh]
\begin{center}
\includegraphics[width=0.9\textwidth,height=0.7\textheight]{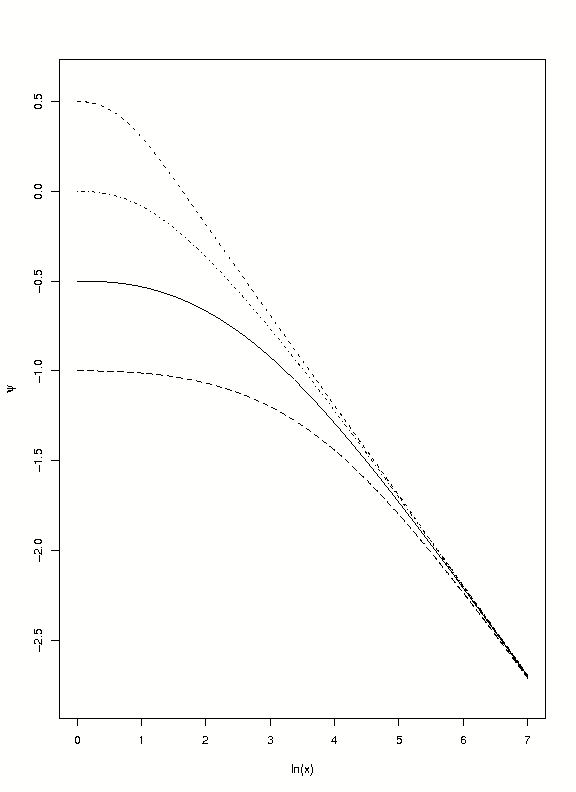}
\end{center}
\caption{A plot of four solutions of equation (\protect\ref{ref3A}) for the
various choices $\protect\psi (\ln x=0)\in \{-1,-0.5,0,0.5\}$ and $d\protect%
\psi /d(\ln x)|_{\ln x=0}=0$, where, again, $x$ is defined as $\ln (t/t_{1})$%
. The parameter values are now $\protect\mu =-4$ and $C=-0.1$. In this case $%
(\protect\mu <-2)$, the solutions eventually converge to a common decreasing
asymptote, and this behaviour is insensitive to the initial conditions.}
\label{fig:3}
\end{figure}


\begin{figure}[tbh]
\begin{center}
\includegraphics[width=0.9\textwidth,height=0.7\textheight]{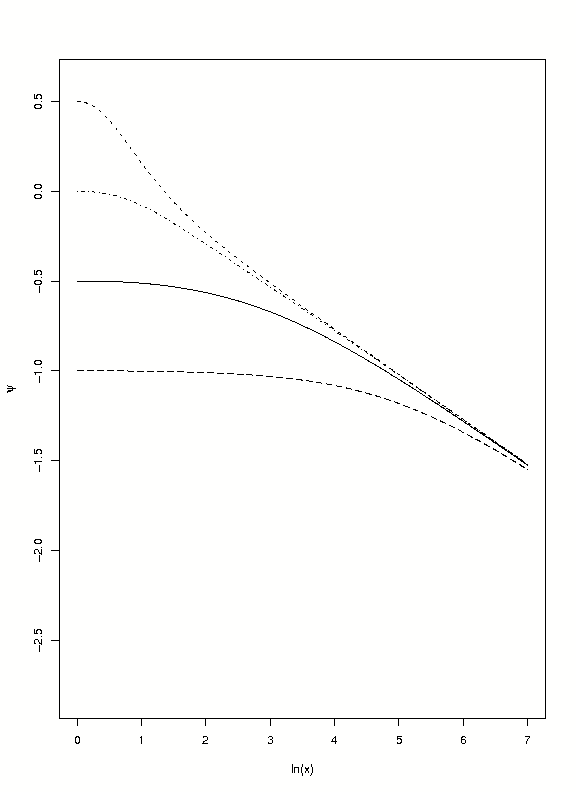}
\end{center}
\caption{As in Figure \protect\ref{fig:3}, but with $\protect\mu =-6$. All
solutions still converge to a common decreasing asymptote at late times.
However, the slope of the asymptote is more gentle than that in Figure 
\protect\ref{fig:3} (because $\protect\mu $ is now more negative). Again,
this slope can be made as gentle as desired by taking a sufficiently large
negative $\protect\mu $.}
\label{fig:4}
\end{figure}

{\bfseries{Summary}}

To summarize, there are three interesting conclusions to be drawn, assuming
this form of $\omega (\psi )=\omega _{0}\exp (\mu \psi )$:

(a) if $\alpha $ decreases during the late matter era, we must have $\mu <-2$
and $C<0$. Note that care must be taken in this case to ensure that the
consistency conditions for the Friedmann equation are not violated, since $%
e^{-2\psi }=[(2+\mu )C\ln t]^{-\frac{2}{2+\mu }}$ is increasing and, if $%
-4<\mu <-2$, then $e^{\mu \psi }\dot{\psi}^{2}t^{2}$ is increasing.

(b) if $\alpha $ increases during the late matter era, we must have $\mu >-2$
and $C>0$. In this case both consistency conditions are automatically
satisfied. This includes the original case of the BSBM theory, in which $\mu
=0$.

(c) it is possible to have special behaviour of $\alpha$ in the case $\mu =
- 2$, and various conditions on $C$ are required depending on whether $%
\alpha $ is decreasing or increasing.

\begin{center}
\textbf{V. SOLUTIONS WHEN $a(t) = a_{0}t^{n}$}
\end{center}

We now consider the evolution of $\alpha (t)$ in situations where the
cosmological scale factor exhibits general power-law behaviour $a=a_{0}t^{n}$%
, where $n>0$. The equation of motion for the scalar field becomes 
\begin{equation}
\ddot{\psi}+\frac{\mu }{2}\dot{\psi}^{2}+\frac{3n}{t}\dot{\psi}=\frac{%
Ce^{-(2+\mu )\psi }}{t^{3n}}
\end{equation}%
where, as before, $\omega =\omega _{0}e^{\mu \psi }$ and $C=N/(\omega
_{0}a_{0}^{3})$. The substitution $x=\ln (t)$ then yields the equation 
\begin{equation}
\psi ^{\prime \prime }+(3n-1)\psi ^{\prime }+\frac{\mu }{2}\psi ^{\prime
2}=Ce^{-(2+\mu )\psi +(2-3n)x}
\end{equation}%
for $\psi (x)$. Two separate cases then arise, depending on whether $\mu =-2$
or $\mu \neq -2$.

{\bfseries{Case 1:}} In the case $\mu =-2$, the evolution equation becomes 
\begin{equation}
\ddot{\psi}-\dot{\psi}^{2}+\frac{3n}{t}\dot{\psi}=\frac{C}{t^{3n}}.
\label{psi}
\end{equation}%
Let $y=\dot{\psi}$. Then 
\begin{equation*}
\dot{y}=\frac{C}{t^{3n}}-\frac{3n}{t}y+y^{2}.
\end{equation*}%
This is a Riccati equation, which can be solved by making the substitution $%
y=-\frac{\dot{u}}{u}$. Then $u$ satisfies 
\begin{equation*}
\text{$\ddot{u}+\frac{3n}{t}\dot{u}+\frac{C}{t^{3n}}u=0$.}
\end{equation*}%
The solution of this differential equation is 
\begin{equation*}
\text{$u=\beta _{1}t^{\frac{1}{2}\left( 1-3n\right) }J_{\frac{1-3n}{3n-2}%
}\left( \frac{2\sqrt{C}(t^{1-\frac{3n}{2}})}{2-3n}\right) +\beta _{2}t^{%
\frac{1}{2}\left( 1-3n\right) }J_{\frac{3n-1}{3n-2}}\left( \frac{2\sqrt{C}%
(t^{1-\frac{3n}{2}})}{2-3n}\right) $}
\end{equation*}%
where $\beta _{1}$ and $\beta _{2}$ are arbitrary constants, and $J_{m}(x)$
is a Bessel function of the first kind. There are two types of late-time
behaviour. If $n<\frac{2}{3}$ then 
\begin{eqnarray*}
u &\approx &\beta _{1}t^{-\frac{3n}{2}}\cos \left( \frac{2\sqrt{C}t^{1-\frac{%
3n}{2}}}{2-3n}-\left( \frac{1-3n}{3n-2}\right) \frac{\pi }{2}-\frac{\pi }{4}%
\right) \\
&&+\beta _{2}t^{-\frac{3n}{2}}\cos \left( \frac{2\sqrt{C}t^{1-\frac{3n}{2}}}{%
2-3n}-\left( \frac{3n-1}{3n-2}\right) \frac{\pi }{2}-\frac{\pi }{4}\right)
\end{eqnarray*}%
so, at late times, $\psi \rightarrow \text{const.}+\frac{3n}{2}\ln t$ and $%
\alpha \rightarrow t^{3n}$. As noted earlier, neither $\psi $ nor $\alpha $
can exhibit oscillatory behaviour because, when $\dot{\psi}$ is zero in (\ref%
{psi}), $\ddot{\psi}$ can only have one sign (determined by the sign of $C$%
). Thus, $\psi $ cannot have both maxima and minima.

If $n>\frac{2}{3}$, then 
\begin{eqnarray*}
u &\approx &\beta _{1}t^{\frac{1-3n}{2}}\left( \frac{\sqrt{C}t^{\frac{2-3n}{2%
}}}{2-3n}\right) ^{\frac{1-3n}{3n-2}}+\beta _{2}t^{\frac{1-3n}{2}}\left( 
\frac{\sqrt{C}t^{\frac{2-3n}{2}}}{2-3n}\right) ^{\frac{3n-1}{3n-2}} \\
&=&A_{1}+A_{2}t^{1-3n}
\end{eqnarray*}%
where $A_{1}$ and $A_{2}$ are new arbitrary constants. Therefore, at late
times, 
\begin{equation*}
\psi \rightarrow \text{const.}-\ln \left( \alpha +\beta t^{1-3n}\right)
\rightarrow \text{const.}
\end{equation*}%
and $\alpha (t)\rightarrow $ constant. Hence, $\psi $ increases as $\ln t$
if $n<\frac{2}{3}$, but tends to a constant if $n>\frac{2}{3}$. The latter
case includes the important scenario of a curvature-dominated Friedmann
universe with $n=1$ and is indicative of the asymptotically constant
behaviour of $\psi $ and $\alpha $ also to be expected in the
dark-energy-dominated situation with $n\rightarrow \infty $. We will discuss
this scenario in Section VI.

{\bfseries{Case 2:}} In the case $\mu \neq -2$, we can define new variables
by $w=e^{-(2+\mu )\psi +(2-3n)x}$ and $z=\psi ^{\prime }(x)$. This leads to
the two-dimensional dynamical system 
\begin{eqnarray*}
w^{\prime } &=&(2-3n)w-(2+\mu )zw \\
z^{\prime } &=&-\frac{\mu }{2}z^{2}+(1-3n)z+Cw.
\end{eqnarray*}%
which has three fixed points: $(w,z)=(0,0),(0,\frac{2}{\mu }(1-3n))$ and $%
\left( \frac{(2-3n)(3n\mu +12n-4)}{2C(2+\mu )^{2}},\frac{2-3n}{2+\mu }%
\right) $. We will call these points $a$, $b$ and $c$ respectively.

Note that the fixed points $a$ and $b$ are asymptotic fixed points, since,
by definition, $w$ never actually reaches zero. On the other hand, point $c$
exists only if $\frac{(2-3n)(3n\mu +12n-4)}{C}>0$. This fixed point
corresponds to the exact solution 
\begin{equation*}
\psi =\frac{1}{2+\mu }\ln \left[ \frac{2C(\mu +2)^{2}}{(2-3n)(3n(\mu +4)-4)}%
\right] +\frac{2-3n}{\mu +2}\ln t,
\end{equation*}%
and this solution reduces to the one found in \cite{BSBM3} for the BSBM ($%
\mu =0$) case. We now investigate the stability of these points in order to
understand the late-time behaviour.

{\bfseries{Critical point $a$.}} For $(0,0),$ the Jacobian matrix has the
eigenvalues $(2-3n)$ and $(1-3n)$. Therefore, if $n\neq \frac{1}{3}$ and $%
n\neq \frac{2}{3}$, this critical point is hyperbolic. It is a saddle point
if $\frac{1}{3}<n<\frac{2}{3}$, an unstable node if $n<\frac{1}{3}$, and a
stable node if $n>\frac{2}{3}$. The case $n<\frac{1}{3}$ is not of physical interest because
it derives from background expansion with $p>\rho $.

{\bfseries{Critical point $b$.}} For $\left( 0, \frac{2}{\mu} (1 - 3 n)
\right)$, the two eigenvalues of the Jacobian are $(3 n - 1)$ and $\left( 
\frac{3 n \mu + 12 n - 4}{\mu} \right)$, so the trace of the Jacobian is $T
= 6 n - 1 + \frac{4}{\mu} (3 n - 1)$ and the determinant is $D = (3 n - 1)
\left( 3 n + \frac{4}{\mu} (3 n - 1) \right)$. It is helpful to note that $%
T^2 - 4 D = \left( \frac{4}{\mu} (3 n - 1) + 1 \right)^2 > 0$ if $\mu \neq 4
(1 - 3 n)$. This means that, in general, the fixed point is either a saddle
point (if $D < 0$) or a node (if $T^2 > 4 D > 0$).

If $n \neq \frac{1}{3}$ and $n \neq \frac{4}{3 (\mu + 4)}$, this critical
point is hyperbolic. Various types of behaviour can occur depending on the
values of $n$ and $\mu$:

\begin{itemizedot}
\item $n>\frac{1}{3}$. Then this point is a saddle if $3n<\frac{4-12n}{\mu }$%
, and an unstable node if $3n>\frac{4-12n}{\mu }$.

\item $n<\frac{1}{3}$: Then this point is a saddle if $3n>\frac{4-12n}{\mu }$%
, and a stable node if $3n<\frac{4-12n}{\mu }$.
\end{itemizedot}

{\bfseries{Critical point $c$.}} In this case, the Jacobian matrix is 
\begin{equation*}
\left( 
\begin{array}{cc}
0 & \frac{(3n-2)(3n\mu +12n-4)}{2(2+\mu )C} \\ 
C & \frac{2-6n-\mu }{2+\mu }%
\end{array}%
\right)
\end{equation*}%
so $T=\frac{2-6n-\mu }{2+\mu }$ and $D=\frac{(2-3n)(3n\mu +12n-4)}{2(2+\mu )}
$. As before, there are several cases. Note that we do not distinguish here
between whether the point is a node or a focus, since our primary interest
is whether the critical point is stable or unstable.

\begin{itemizedot}
\item If $n > \frac{2}{3}$ and $\mu < \frac{4}{3 n} - 4$, point $c$ is a
saddle point.

If $n > \frac{2}{3}$ and $\frac{4}{3 n} - 4 < \mu < - 2$, point $c$ is an
unstable node/focus.

If $n > \frac{2}{3}$ and $\mu > - 2$, point $c$ is a saddle point.

\item If $n < \frac{2}{3}$ and $\mu < - 2$, point $c$ is a stable node/focus.

If $n < \frac{2}{3}$ and $- 2 < \mu < \frac{4}{3 n} - 4$, point $c$ is a
saddle point.

If $n < \frac{2}{3}$ and $\mu > \frac{4}{3 n} - 4$, then point $c$ is a
stable node/focus if $\mu > 2 - 6 n$, and is an unstable node/focus if $\mu
< 2 - 6 n$. (Note that this means that if $n < \frac{1}{3}$ and $\mu > \frac{%
4}{3 n} - 4$, point $c$ must be a stable node/focus.)
\end{itemizedot}

We can use these results to obtain a general result for the case $n>\frac{2}{%
3}$ by noting that the only stable critical point is at $(w,z)=(0,0)$.
Therefore, assuming that $z=t\dot{\psi}$ does not diverge to infinity, we
are likely to obtain the behaviour $\psi \rightarrow \text{const.}$ at late
times in accelerating universes.

\begin{center}
\textbf{VI. SOLUTIONS IN AN ACCELERATING UNIVERSE}
\end{center}

The data from the Oklo natural reactor ($z\approx 0.14$) {\cite{OKLO}}, the
analysis of meteorites ($z\approx 0.45$) {\cite{METEOR}} and laboratory
experiments \cite{lab} provide very tight bounds on the allowed variation of
alpha at low redshifts. We will show that our generalization of the BSBM
theory is still consistent with negligible variation in $\alpha $ during the
accelerating era of the universe, although we do not introduce the added
complexity of relating local terrestrial observations of varying constants
to their global cosmological evolution (see Barrow and Shaw \cite%
{BarrowShaw1}, \cite{BarrowShaw2}, \cite{bshaw} for a detailed discussion of
this problem).

The cases of power-law expansion with $n>2/3$ discussed in the last section
cover the situation of an accelerating universe dominated by a fluid with
equation of state $-\rho <p<-\rho /3$. The remaining situation of special
interest is the particular case of a $\Lambda $-dominated era, with $p=-\rho 
$, in which the scale factor evolves according to the de Sitter form%
\begin{equation*}
a=a_{0}e^{Mt},
\end{equation*}%
where $M\equiv \sqrt{\Lambda /3}$. The evolution equation (\ref{evol}) for $%
\psi $ becomes 
\begin{equation}
\ddot{\psi}+\frac{\mu }{2}\dot{\psi}^{2}+3M\dot{\psi}=\frac{Ce^{-(2+\mu
)\psi }}{e^{3Mt}}  \label{4A}
\end{equation}%
where $C=N/(\omega _{0}a_{0}^{3})$, and the consistency conditions (\ref%
{consist}) are $e^{-2\psi }\ll 1$ and $\dot{\psi}^{2}e^{\mu \psi }\ll 1$.

\begin{figure}[tbh]
\begin{center}
\includegraphics[width=0.9\textwidth,height=0.7\textheight]{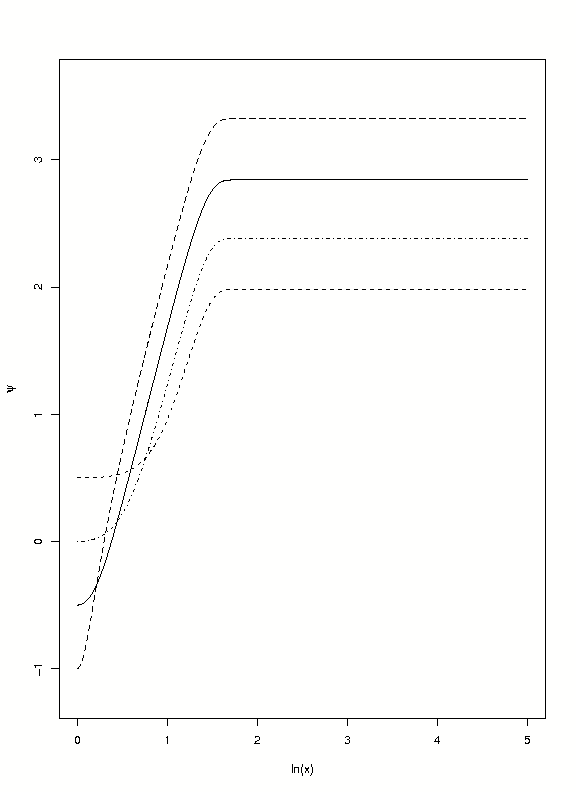}
\end{center}
\caption{As in Figure \protect\ref{fig:1}, but these solutions now occur
within an accelerating universe. The parameter values are now $\protect\mu %
=2 $, $C=0.1$ and $M=0.01$. Regardless of the initial conditions, the
solutions all tend to constant values after a short initial time. This
illustrates how the time variation of $\protect\alpha $ is suppressed in a $%
\Lambda $-dominated universe.}
\label{fig:5}
\end{figure}

Numerical investigation shows that there are two types of solution. One type
tends to a constant at late times, in general increasing before it does so.
The value of $\psi $ along the eventual constant asymptote depends on the
initial conditions. A notable feature of Fig. \ref{fig:5} is that the
behaviour of $\psi $ during the initial stages of the evolution can be
complicated, but all solutions seem to become constant for $t>t^{\ast }$,
where $t^{\ast }$ is some fixed `cooling-off' time. Therefore, if the matter
era lasts past time $t^{\ast }$, we would expect negligible variation in $%
\psi $ once the cosmological constant begins to dominate.

We might expect this behaviour from (\ref{4A}) if $\dot{\psi}$ is not too
large, because then we have, approximately, 
\begin{equation*}
\ddot{\psi}=-3M\dot{\psi}.
\end{equation*}%
Since $3M>0$, $\dot{\psi}$ is driven exponentially quickly to zero, and $%
\psi $ tends to a constant.

The second case occurs when $\dot{\psi}$ is large, so that the $\frac{\mu }{2%
}\dot{\psi}^{2}$ term comes into play. We then have 
\begin{equation}
\ddot{\psi}=-\left( \frac{\mu }{2}\dot{\psi}+3M\right) \dot{\psi},
\end{equation}%
and there are two equilibria for $\dot{\psi}$: at $\dot{\psi}=0$ and $\dot{%
\psi}=-6M/\mu $. Regardless of the sign of $\mu $, the fixed point at $\dot{%
\psi}=0$ is stable and the fixed point at $\dot{\psi}=-\frac{6M}{\mu }$ is
unstable. Therefore, depending on the initial conditions, there are two
generic behaviours for $\dot{\psi}$: it may either tend to $\pm \infty $
exponentially quickly, or tend to zero. Assuming that finite-time blow-up
does not occur, $\dot{\psi}$ should tend to zero at late times, so $\psi $
tends to a constant as found above. Otherwise, we obtain a second class of
diverging solutions.

We can also perform an analysis similar to the one performed in Section V
for the general power-law case, but this time we will work directly with
equation (\ref{4A}). We begin by defining the variables $w=e^{-(2+\mu )\psi
-3Mt}$ and $z=\dot{\psi}$. This leads to the autonomous dynamical system 
\begin{eqnarray*}
\dot{w} &=&-3Mw-(2+\mu )zw, \\
\dot{z} &=&-\frac{\mu }{2}z^{2}-3Mz+Cw.
\end{eqnarray*}%
Again, this system has three fixed points, at $(w,z)=(0,0),(0,-\frac{6M}{\mu 
})$ and $\left( -\frac{9M^{2}(4+\mu )}{2C(2+\mu )^{2}},-\frac{3M}{2+\mu }%
\right) $. As before, we will name these points $a$, $b$ and $c$
respectively. Again, points $a$ and $b$ are asymptotic fixed points, and
point $c$ only exists if $(4+\mu )/C<0$. The analysis of the stability of
these fixed points now follows:

{\bfseries{Critical point $a$.}} At $(0, 0)$, the Jacobian has the two
eigenvalues $- 3 M$. Therefore, the origin is always a stable node.

{\bfseries{Critical point $b$.}} At $\left( 0, - \frac{6 M}{\mu} \right)$,
the Jacobian has eigenvalues $3 M \left( 1 + \frac{4}{\mu} \right)$ and $3 M$%
. The trace and determinant are therefore given by $T = 6 M \left( \frac{2}{%
\mu} + 1 \right)$ and $D = 9 M^2 \left( \frac{4}{\mu} + 1 \right)$
respectively, and we have $T^2 - 4 D = 144 > 0$. Hence, if $- 4 < \mu < 0$,
this critical point is a saddle. Otherwise, if $\mu > 0$ or $\mu < - 4$, it
is an unstable node.

{\bfseries{Critical point $c$.}} If this fixed point exists, the
corresponding Jacobian matrix has determinant $D = - \frac{9 M^2 (4 + \mu)}{%
2 (2 + \mu)}$ and trace $T = - \frac{6 M}{2 + \mu}$. Therefore, if $- 4 <
\mu < - 2$, this point is an unstable node, and if either $\mu > - 2$ or $%
\mu < - 4$, it is a saddle point.

Note that, for any value of $M$, the origin $(0,0)$ is the only stable node.
Therefore, in an accelerating era, any bounded late-time solution will
satisfy $\psi \rightarrow \text{const.}$, and we have a form of the cosmic
no-hair theorem for the evolution of $\psi $ and $\alpha $ in the de Sitter
background metric.

\begin{center}
\textbf{VII. SUMMARY AND THE RELATION TO OBSERVATIONS}
\end{center}

We are now in a position to summarize the results of the previous sections,
having determined the late-time behaviors for both power-law and exponential
evolution of the scale factor $a(t)$. For convenience, the different
outcomes are displayed in Table 4.1.

\begin{table}[h]
\resizebox{13cm}{!} {

\renewcommand{\arraystretch}{2.3}
\begin{tabular}{|c|l|l|c|c|}
\hline
\multirow{2}{*}{Value of $\mu$} & \multicolumn{3}{c|}{Scale factor evolution $a(t) \propto t^n$} & \multirow{2}{*}{\text{   }$a(t) \propto e^{M t} \text{  } $} \\
\cline{2-4}
                                & \multicolumn{1}{c}{$\frac{1}{3} < n < \frac{2}{3}$} & \multicolumn{1}{|c}{$n = \frac{2}{3}$} & \multicolumn{1}{|c|}{$n > \frac{2}{3}$} &  \\
\hline
$\mu > 0$ & $\alpha \rightarrow t^{\frac{2 (2 - 3 n)}{2 + \mu}}$ \ \ if $C > 0$  & $\alpha \rightarrow \left[ (2 + \mu) C \ln (t / t_1) \right]^{\frac{2}{2 + \mu}}$ \ if $C > 0$ & const & const\\
\hline
$\mu = 0$ (BSBM) & $\alpha \rightarrow t^{2 - 3 n}$ \ \ \ \ if $C > 0$ & $\alpha \rightarrow 2 C \ln (t / t_1)$ \ \ \ \ \ \ \ \ \ \ \ \ \ \ if $C > 0$ & const & const\\ 
\hline
$- 2 < \mu < 0$ & $\alpha \rightarrow t^{\frac{2 (2 - 3 n)}{2 + \mu}}$\  if $C > 0$ & $\alpha \rightarrow \left[ (2 + \mu) C \ln (t / t_1) \right]^{\frac{2}{2 + \mu}}$ if $C > 0$ & const & const\\
\hline
$\mu = - 2$ & $\alpha \rightarrow t^{3 n}$ \ \ \ \ \ \  if $C > 0$ & various cases, depending on $C$ & const & const\\
\hline
$\mu < - 2$ & $\alpha \rightarrow t^{\frac{2 (2 - 3 n)}{2 + \mu}}$ if $C < 0$ & $\alpha \rightarrow \left[ (2 + \mu) C \ln (t / t_1) \right]^{\frac{2}{2 + \mu}}$ if $C < 0$ & const & const\\
\hline
\end{tabular}
}
\caption{This table lists the late-time behaviours of $\protect\alpha (t)$
for various choices of $\protect\mu $ and possible evolutions of the scale
factor $a(t)$, of both power-law and exponential forms. In the power-law
case with $n>2/3$, and in the exponential case, the fine structure
`constant' $\protect\alpha (t)$ tends to a constant value at late times. In
other cases, $\protect\alpha (t)$ tracks an increasing asymptote if $\protect%
\mu >-2$, and a decreasing asymptote if $\protect\mu <-2$.}
\end{table}

We can now attempt to constrain the parameters of the theory by using
observational data \cite{DATA}. Since most of the observational evidence
arises from the late matter-dominated era, we will use the asymptotic
behaviour obtained in earlier sections as the theoretical basis for
comparison.

In the matter-dominated era, the redshift-time relation 
\begin{equation*}
\frac{a(t_{1})}{a(t_{2})}=\frac{1+z_{2}}{1+z_{1}}=\left( \frac{t_{1}}{t_{2}}%
\right) ^{2/3}
\end{equation*}%
holds. From our results in Section IV above, solutions tend to the asymptote 
\begin{equation*}
\psi \rightarrow \frac{1}{2+\mu }\ln \left( (2+\mu )C\ln (t/t_{1})\right)
\end{equation*}%
where $t_{1}$ is an arbitrary constant (corresponding to redshift $z_{1}$,
say). Then, using (\ref{alpha}), we can express the relative shift in the
value of the fine-structure constant expected between a redshift $z\gtrsim
z_{\Lambda }\sim 0.5$ and the present by 
\begin{eqnarray}
\frac{\Delta \alpha }{\alpha } &\rightarrow &\frac{\alpha (z)-\alpha _{0}}{%
\alpha _{0}}=\frac{\alpha (z)-\alpha _{\Lambda }}{\alpha _{\Lambda }}  \notag
\\
&=&\left( \frac{\ln (1+z_{1})-\ln (1+z)}{\ln (1+z_{1})-\ln (1+z_{\Lambda })}%
\right) ^{\frac{2}{2+\mu }}-1  \notag
\end{eqnarray}%
where $\alpha _{\Lambda }=\alpha _{0}$ (to an excellent approximation) is
the value of $\alpha $ at the redshift signifying the onset of $\Lambda $%
-domination and an accelerating universe.

Thus, defining the new parameters $\nu =2/(2+\mu )$ and $Z=\ln (1+z_{1})$,
we obtain the following approximate form for the evolution of $\Delta \alpha
/\alpha $ during the matter era: 
\begin{equation}
\frac{\Delta \alpha }{\alpha }\rightarrow \left( \frac{Z-\ln (1+z)}{Z-\ln
(1+z_{\Lambda })}\right) ^{\nu }-1.  \label{gen}
\end{equation}%
The only constraint on $Z$ is that it should correspond to a redshift before
dust domination, i.e., $Z\gtrsim 10$. Taking $z_{\Lambda }=0.5,$ sampling
values of $Z$ between 0.1 and 1000, and values of $\nu $ between $-100$ and $%
100$, indicates that the best-fit parameters generally satisfy the
approximate relation $\nu \sim 2.5Z$, or $(2+\mu )\ln (1+z_{1})\sim 0.8$.
The associated $\chi ^{2}$ value is approximately $168.75$, with 143 degrees
of freedom; this indicates a fit that is reasonable but not particularly
strong. Formula (\ref{gen}) generalises the analytic approximation for $%
\Delta \alpha /\alpha $ obtained in the BSBM theory ($\mu =0$).

In the above, we have assumed a sharp transition between dust domination and
de Sitter expansion at $z=z_{\Lambda }$, and assumed that $\psi $ and $%
\alpha $ are constant when $z<z_{\Lambda }$. A more accurate approximation
that evolves smoothly between the dust and $\Lambda $-dominated eras could
be obtained by assuming a background cosmological evolution in (\ref{evol})
where 
\begin{equation}
a^{3}=\frac{8\pi G\rho _{0}a_{0}^{3}}{\Lambda }\sinh ^{2}\left( \frac{\sqrt{%
3\Lambda }}{2}t+\beta \right) .
\end{equation}%
This form of $a(t)$ solves the Friedmann equation 
\begin{equation*}
\frac{\dot{a}^{2}}{a^{2}}=\frac{8\pi G}{3}\rho _{m}+\frac{\Lambda }{3}
\end{equation*}%
for a universe containing dust and a cosmological constant. In this model,
the fractional densities are $\Omega _{m}=\text{sech}^{2}\left( \frac{\sqrt{%
3\Lambda }}{2}t\right) $ and $\Omega _{\Lambda }=\tanh ^{2}\left( \frac{%
\sqrt{3\Lambda }}{2}t\right) $ respectively. Using the approximate values $%
\Omega _{\Lambda 0}=0.73$ and $\Omega _{m0}=0.27$ \cite{WMAP} and $H_{0}=74.2%
\text{km}\text{s}^{-1}\text{Mpc}^{-1}$ \cite{RIESS}, we obtain $\Lambda
=1.266340\times 10^{-35}\text{s}^{-2}$ and $8\pi G\rho
_{0}=0.81H_{0}^{2}=4.684\times 10^{-36}\text{s}^{-2}$. We have data for 
\begin{equation*}
\frac{\Delta \alpha }{\alpha }=e^{2(\psi -\psi _{0})}-1\equiv e^{-2\psi
_{0}}f-1
\end{equation*}%
and we are interested in the evolution of $f(z)\equiv e^{2\psi }$.

\begin{figure}[tbh]
\begin{center}
\includegraphics[width=0.7\textwidth,height=0.8%
\textheight]{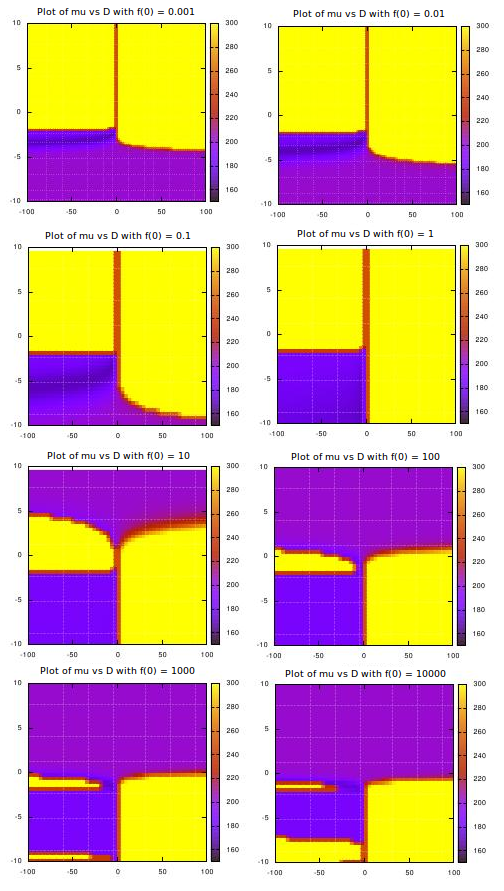}
\end{center}
\caption{These contour plots display the value of the $\protect\chi ^{2}$
statistic (according to the legend shown in the right-hand color bars) for
various choices of $D$, $\protect\mu $ and $\protect\psi _{0}$ in a model
that exhibits a smooth transition between the dark-matter and dark-energy
dominated eras. Each picture corresponds to a different value of $f(0)\equiv
e^{2\protect\psi (0)}$ ranging from $10^{-3}$ to $10^{4}$. The vertical and
horizontal axes represent the parameters $\protect\mu $ and $D$,
respectively. Darker regions in the plots indicate parameter choices that
give rise to a lower $\protect\chi ^{2}$ statistic, and therefore correspond
to a better fit.}
\label{fig:6}
\end{figure}

The evolution equation for $\psi (t),$ ($\ref{ref3A}$), can be transformed
into an evolution equation for $f(z)$: 
\begin{eqnarray*}
&&\left[ 8\pi G\rho _{0}(1+z)^{3}+\Lambda \right] (1+z)\left[ f^{\prime
\prime }+\frac{\mu f^{\prime 2}}{4f}-\frac{f^{\prime 2}}{f}\right] -\left[
4\pi G\rho _{0}(1+z)^{3}+2\Lambda \right] f^{\prime } \\
&=&48\pi GD\rho _{0}(1+z)^{2}f^{-\frac{\mu }{2}},
\end{eqnarray*}%
where $D=-\frac{\zeta }{4\pi G\omega _{0}}$. This allows us to determine the
behaviour of $f$ for various initial conditions $f(0)$ and $f^{\prime }(0)$.
Note that $f(0)=e^{2\psi _{0}}>0$, and, since the present-day variation of $%
\alpha $ is negligible, we have assumed that $f^{\prime }(0)=0$. Thus, our
system has three arbitrary parameters: the value of $f(0)$, the value of $%
\mu $, and the value of $D$. The contour plots in Fig. \ref{fig:6}
illustrate the regions of parameter space for which there is a good fit to
the observed data. The best chi-squared statistic obtained using this model
is approximately $169$ (attained when $f(0)<1$), which still indicates a fit
that is not very strong.

It is important to note that these investigations of the time variation of $\alpha$ 
may not be conclusive because recent observations of $\alpha$ are consistent with a
spatial variation, and the latter will need to be accounted for before an accurate estimate 
of $\mu $ (or an accurate test of whether $\mu \neq 0$) can be obtained. If we
are interested in modelling possible spatial variations in $\alpha $
at the redshifts of quasar absorption spectra, the BSBM theory would require
inhomogeneity in $\zeta /\omega $. Since $\omega $ is constant in that
theory, it would be necessary for $\zeta $ to vary in space. This means that
the identity of the dominant form of cold dark matter would vary throughout
space, or has a strange character that requires spatial variation in its
electric and magnetic composition. This is not appealing in the absence of
other evidence because the time evolutions of the densities of these
different varieties of pressureless matter will be the same even if they
have different $\zeta $ values, and, for one value of $\zeta $ to be
replaced by another, some type of decay would need to occur. However, in the
generalized theory explored here, inhomogeneity in $\alpha $ can be caused
by spatial variation in $\zeta /\omega (\psi )$ via $\psi (\vec{x})$, and
this is not unnatural. This situation will be explored in greater detail
elsewhere by extending the results of \cite{vary} and \cite{pert}.

\begin{center}
\textbf{\ }   

\textbf{VIII. CONCLUSIONS}
\end{center}

We have shown that the BSBM theory of varying alpha naturally generalises to
cases in which the scalar field coupling $\omega (\psi )>0$ has a functional
dependence on the scalar field $\psi$. We have explored the features of the
specific case where this dependence is exponential, $\omega (\psi )\propto
e^{\mu \psi }$. The BSBM solution, which is recovered when $\mu =0$, may now
be viewed as just one of a range of possible solutions with $\mu >-2$ which
all exhibit similar asymptotic cosmological behaviour in dust or
dark-energy-dominated eras at late times. In all these cases, $\alpha $
increases slowly with time, or (in the dark-energy-dominated cases)
increases and then tends to a constant-valued asymptote. In the case of a de
Sitter background, this gives an extension of the cosmic no-hair behaviour
familiar from studies of general relativistic cosmology with constant $%
\alpha .$

We have also found different possible behaviors in the cases $\mu \leq -2$.
In these cases, it is possible for $\psi $ (and $\alpha $) to decrease with
time, and the consistency conditions must be checked numerically in order to
ensure that our approximate solution to the Friedmann equation is not
changed by the growing gravitational effects of decreasing $\psi $. 

Finally, it is important to note the limitations of this analysis. In this paper, we
have confined attention to theories where the late-time behaviour of $a(t)$
is unaffected by the varying scalar field. However, if this was not the
case, then new possibilities might arise. We have also highlighted the need for an
analysis of the effects of spatial variation of $\psi $ and $\alpha $ in 
the post-recombination era; this work will be presented elsewhere.

\begin{center}
\textbf{ACKNOWLEDGMENTS}
\end{center}

S. Lip would like to thank the Gates Cambridge Trust for its support. We
also thank Anne Davis, Jo\~{a}o Magueijo, Michael Murphy and John Webb for
discussions and Sergei Levshakov for helpful comments..

\end{document}